\magnification=1100
\hsize=6.5 true in
\vsize=9.0 true in
\overfullrule=0pt
\baselineskip=15pt
\tolerance=10000
\hbadness=200

\baselineskip 15.5pt
\font\tensc=cmcsc10
\newfam\scfam
\textfont\scfam=\tensc
\def\sc{\fam\scfam\tensc}
 
\font\bigbf=cmbx10 scaled\magstep1

\font\biggbf=cmbx10 scaled\magstep2

\font\bigggbf=cmbx10 scaled\magstep3

\font\bigit=cmti10 scaled\magstep1

\def\sectionbreak{
\bigskip\vskip\parskip}

\def\bigsectionbreak{
\bigskip\bigskip\vskip\parskip}

\outer\def\beginsection#1\par
{\sectionbreak
\message{#1}\leftline{\bf#1}\nobreak\smallskip\noindent}

\outer\def\bigbeginsection#1\par
{\sectionbreak
\message{#1}\leftline{\bigbf#1}\nobreak\medskip\noindent}

\def\currentsection{\firstmark}

\outer\def\biggbeginsection#1\par
{\bigsectionbreak
\message{#1}\leftline{\biggbf#1}
\mark{#1}\nobreak\bigskip\noindent}

\outer\def\bigitbeginsection#1\par
{\sectionbreak
\message{#1}\leftline{\bigit#1}\nobreak\medskip\noindent}

\outer\def\longbigbeginsection#1 #2\par#3\par
{\sectionbreak
\message{#1 #2 #3}
\halign{##\hfil&##\hfil\cr
{\bigbf#1\ }&{\bigbf#2}\cr
&{\bigbf#3}\cr
}\nobreak\medskip\noindent}

\outer\def\longbiggbeginsection#1 #2\par#3\par
{\bigsectionbreak
\message{#1 #2 #3}
\halign{##\hfil&##\hfil\cr
{\biggbf#1\ }&{\biggbf#2}\cr
&{\biggbf#3}\cr
}\mark{#1 #2 #3}\nobreak\bigskip\noindent}

\outer\def\longbiggbeginappendix#1 #2 #3\par#4\par
{\bigsectionbreak
\message{#1 #2 #3 #4}
\halign{##\hfil&##\hfil\cr
{\biggbf#1\ #2\ }&{\biggbf#3}\cr
&{\biggbf#4}\cr
}\mark{#1 #2 #3 #4}\nobreak\bigskip\noindent}

\def\rightheadline{\hfil{\it\currentsection}\ \ \ \ \ \ {\rm \folio}}

\def\currentchapter{}

\def\leftheadline{{\rm \folio}\ \ \ \ \ \ {\it\currentchapter}\hfil}

\newcount\titlepageno

\def\setheadline{\headline=
{\ifnum\titlepageno=\pageno{\hfil}
\else{\ifodd\pageno{\rightheadline}\else{\leftheadline}\fi}
\fi}}

\def\skipifeven
{\ifodd\pageno{}\else\advancepageno\fi}

\outer\def\beginchapter#1. #2\par
{\vfill\eject\skipifeven\titlepageno=\pageno
\def\currentchapter{Cap\'\i tulo #1. #2}
\topinsert\vskip 0.25\vsize\endinsert
\hrule\medskip
\rightline{\bigggbf Cap\'\i tulo #1}
\medskip
\rightline{\bigggbf#2}
\medskip\hrule\bigskip\bigskip}

\outer\def\shortbeginchapter#1\par
{\vfill\eject\skipifeven\titlepageno=\pageno
\def\currentchapter{#1}
\mark{#1}
\topinsert\vskip 0.25\vsize\endinsert
\hrule\medskip
\rightline{\bigggbf#1}
\medskip\hrule\bigskip\bigskip}

\def\LaTeX{{\rm L\kern-.36em\raise.3ex\hbox{\sc a}\kern-.15em%
    T\kern-.1667em\lower.7ex\hbox{E}\kern-.125emX}}

\def\mytilde{\kern-.5pt\lower3pt\hbox{\char'176}\kern.5pt}

\input epsf
\newdimen\textitemamount\textitemamount=30pt

\centerline{\biggbf On Reducing the Complexity of Matrix Clocks}

\bigskip\bigskip
\centerline{\it L\'ucia M. A. Drummond}

\medskip
\centerline{Universidade Federal Fluminense}
\centerline{Instituto de Computa\c c\~ao}
\centerline{Rua Passo da P\'atria, 156}
\centerline{24210-240 Niter\'oi - RJ, Brazil}
\centerline{\tt lucia@dcc.ic.uff.br}

\bigskip
\centerline{\it Valmir C. Barbosa}

\medskip
\centerline{Universidade Federal do Rio de Janeiro}
\centerline{Programa de Engenharia de Sistemas e Computa\c c\~ao, COPPE}
\centerline{Caixa Postal 68511}
\centerline{21941-972 Rio de Janeiro - RJ, Brazil}
\centerline{\tt valmir@cos.ufrj.br}

\bigskip\bigskip
\centerline{\bf Abstract}

\medskip
\noindent
Matrix clocks are a generalization of the notion of vector clocks that allows
the local representation of causal precedence to reach into an asynchronous
distributed computation's past with depth $x$, where $x\ge 1$ is an integer.
Maintaining matrix clocks correctly in a system of $n$ nodes requires that every
message be accompanied by $O(n^x)$ numbers, which reflects an exponential
dependency of the complexity of matrix clocks upon the desired depth $x$. We
introduce a novel type of matrix clock, one that requires only $nx$ numbers to
be attached to each message while maintaining what for many applications may be
the most significant portion of the information that the original matrix clock
carries. In order to illustrate the new clock's applicability, we demonstrate
its use in the monitoring of certain resource-sharing computations.

\bigskip\bigskip
\noindent
{\bf Keywords:} Vector clocks, matrix clocks.

\vfill
\noindent
{\bf Address for correspondence and proofs:}

\medskip
Valmir C. Barbosa, {\tt valmir@cos.ufrj.br}

Programa de Sistemas, COPPE/UFRJ

Caixa Postal 68511

21941-972 Rio de Janeiro - RJ

Brazil

\eject
\baselineskip 17.60pt
\bigbeginsection 1. Introduction and background

We consider an undirected graph $G$ on $n$ nodes. Each node in $G$ stands for a
computational process and undirected edges in $G$ represent the possibilities
for bidirectional point-to-point communication between pairs of processes. A
fully asynchronous distributed computation carried out by the distributed system
represented by $G$ can be viewed as a set of events occurring at the various
nodes. An event is the sending of a message by a node to another node to which
it is directly connected in $G$ (a neighbor in $G$), or the reception of a
message from such a neighbor, or yet the occurrence at a node of any internal
state transition of relevance (``state'' and ``relevance'' here are highly
dependent upon the particular computation at hand, and are left unspecified).

The standard framework for analyzing such a system is the partial order,
often denoted by $\prec$, that formalizes the usual ``happened-before'' notion
of distributed computing [2, 15]. This partial order is the transitive closure
of the more elementary relation to which the ordered pair $(v,v')$ of events
belongs if either $v$ and $v'$ are consecutive events at a same node in $G$
or $v$ and $v'$ are, respectively, the sending and receiving of a message
between neighbors in $G$.

At node $i$, we let local time be assessed by the number $t_i$ of events that
already occurred. We interchangeably adopt the notations $t_i={\it time}_i(v)$
and $v={\it event}_i(t_i)$ to indicate that $v$ is the $t_i$th event occurring
at node $i$, provided $t_i\ge 1$. Another important partial order on the set of
events is the relation that gives the predecessor at node $j\neq i$ of an event
$v'$ occurring at node $i$. We say that an event $v$ is such a predecessor,
denoted by $v={\it pred}_j(v')$, if $v$ is the event occurring at node $j$ such
that $v\prec v'$ for which ${\it time}_j(v)$ is greatest. If no such $v$ exists,
then ${\it pred}_j(v')$ is undefined and
${\it time}_j\bigl({\it pred}_j(v')\bigr)$ is assumed to be zero.

The relation ${\it pred}_j$ allows the definition of vector clocks [11--14, 17],
as follows. The vector clock of node $i$ at time $t_i$ (that is, following the
occurrence of $t_i$ events at node $i$), denoted by $V^i(t_i)$, is a vector
whose $j$th component, for $1\le j\le n$, is either equal to $0$, if $t_i=0$,
or given by
$$
V^i_j(t_i)=\cases{
t_i,&if $j=i$;\cr
{\it time}_j\Bigl({\it pred}_j\bigl({\it event}_i(t_i)\bigr)\Bigr),
&if $j\neq i$,\cr
}\eqno(1)
$$
if $t_i\ge 1$. In other words, $V^i_j(t_i)$ is either the current time at node
$i$, if $j=i$, or is the time at node $j$ that results from the occurrence at
that node of the predecessor of the $t_i$th event of node $i$, otherwise. If no
such event exists (i.e., $t_i=0$), then $V^i_j(t_i)=0$.

Vector clocks evolve following two simple rules:

{\parindent=\textitemamount

\medskip
\item{$\bullet$} Upon sending a message to one of its neighbors, node $i$
attaches $V^i(t_i)$ to the message, where $t_i$ is assumed to already account
for the message that is being sent.

\medskip
\item{$\bullet$} Upon receiving a message from node $k$ with attached vector
clock $V^k$, node $i$ sets $V^i_i(t_i)$ to $t_i$ (which is assumed to already
reflect the reception of the message) and $V^i_j(t_i)$ to
$\max\bigl\{V^i_j(t_i-1),V^k_j\bigr\}$, for $j\neq i$.

}

\medskip\noindent
It is a simple matter to verify that these rules do indeed maintain vector
clocks consistently with their definition [11--14, 17]. Under these rules or
variations thereof, vector clocks have proven useful in a variety of distributed
algorithms to detect some types of global predicates [10, 14].

For large $n$, attaching a vector clock to every message that is sent is likely
to become burdensome, so the question arises whether less costly implementations
are possible. Under the very general assumptions we have made concerning the
nature of $G$ as a distributed system, the answer is negative: a result similar
to Dilworth's theorem on partially ordered sets [9] establishes that the
size-$n$ attachments are necessary [7]. However, it is possible to use more
economical attachments if the edges of $G$ provide FIFO communication [22], or
if certain aspects of the structure of $G$ can be taken into account [19], or
yet if the full capabilities of vector clocks are not needed [1].

One natural generalization of the notion of vector clocks is the notion of
matrix clocks [21, 24]. For an integer $x\ge 1$, the $x$-dimensional matrix
clock of node $i$ at time $t_i$, denoted by $M^i(t_i)$, has $O(n^x)$ components.
For $1\le j_1,\ldots,j_x\le n$, component $M^i_{j_1,\ldots,j_x}(t_i)$ of
$M^i(t_i)$ is only defined for $i=j_1=\cdots=j_x$ and for
$i\neq j_1\neq\cdots\neq j_x$. As in the definition of $V^i(t_i)$,
$M^i_{j_1,\ldots,j_x}(t_i)=0$ if $t_i=0$. For $t_i\ge 1$, on the other hand,
we have
$$
M^i_{j_1,\ldots,j_x}(t_i)=\cases{
t_i,&if $i=j_1=\cdots=j_x$;\cr
{\it time}_{j_x}\Bigl({\it pred}_{j_x}\ldots{\it pred}_{j_1}
\bigl({\it event}_i(t_i)\bigr)\Bigr),
&if $i\neq j_1\neq\cdots\neq j_x$,\cr
}\eqno(2)
$$
which, for $i\neq j_1\neq\cdots\neq j_x$, first takes the predecessor at node
$j_1$ of the $t_i$th event occurring at node $i$, then the predecessor at node
$j_2$ of that predecessor, and so on through node $j_x$, whose local time after
the occurrence of the last predecessor in the chain is assigned to
$M^i_{j_1,\ldots,j_x}(t_i)$. Should any of these predecessors be undefined, the
ones that follow it in the remaining nodes are undefined as well, and the local
time that results at the end is zero. It is straightforward to see that, for
$x=1$, this definition is equivalent to the definition of a vector clock in (1).
Similarly, the maintenance of matrix clocks follows rules entirely analogous to
those used to maintain vector clocks [14].

While the $j$th component of the vector clock following the occurrence of event
$v'$ at node $i\neq j$ gives the time resulting at node $j$ from the occurrence
of ${\it pred}_j(v')$, the analogous interpretation that exists for matrix
clocks requires the introduction of additional notation. Specifically, the
definition of a set of events encompassing the possible $x$-fold compositions of
the relation ${\it pred}_j$ with itself, denoted by ${\it Pred}^{(x)}_j$, is
necessary. If an event $v$ occurs at node $j$, then we say that
$v\in{\it Pred}^{(x)}_j(v')$ for an event $v'$ occurring at node $i$ if one of
the following holds:

{\parindent=\textitemamount

\medskip
\item{$\bullet$} $x=1$, $j\neq i$, and $v={\it pred}_j(v')$.

\medskip
\item{$\bullet$} $x>1$ and there exists $k\neq i$ such that an event $\bar v$
occurs at node $k$ for which $\bar v={\it pred}_k(v')$ and
$v\in{\it Pred}^{(x-1)}_j(\bar v)$.

}

\medskip\noindent
Note that this definition requires $j\neq k$, in addition to $k\neq i$,
for $v\in{\it Pred}^{(x)}_j(v')$ to hold when $x=2$.

For $x>1$ and $i\neq j_1\neq\cdots\neq j_x=j$, this definition allows the
following interpretation of entry $j_1,\ldots,j_x$ of the matrix clock that
follows the occurrence of event $v'$ at node $i$, that is, of
$M^i_{j_1,\ldots,j_{x-1},j}\bigl({\it time}_i(v')\bigr)$. It gives the time
resulting at node $j$ from the occurrence of an event that is in the set
${\it Pred}^{(x)}_j(v')$, so long as it is nonempty. Of course, the number of
possibilities for such an event is $O(n^{x-1})$ in the worst case, owing to the
several possible combinations of $j_1,\ldots,j_{x-1}$.

Interestingly, applications that require the full capabilities of matrix
clocks have yet to be identified. In fact, it seems a simple matter to argue
that a slightly more sophisticated use of vector clocks or simply the use of
two-dimensional matrix clocks (the $x=2$ case) suffices to tackle some of the
problems that have been offered as possible applications of higher-dimensional
matrix clocks [14, 18]. What we do in this paper is to demonstrate how the
distributed monitoring of certain resource-sharing computations can benefit from
the use of matrix clocks and that, at least for such computations, it is
possible to employ much less complex matrix clocks (that is, matrix clocks with
many fewer components), which nonetheless retain the ability to reach into the
computation's past with arbitrary depth.

The key to this reduction in complexity is the use of one single event in place
of each set ${\it Pred}^{(y)}_j(v')$, for $1\le y\le x$. In other words, for
each node $j$ the matrix clock we introduce retains only one of the
$O(n^{y-1})$ components of each of the $x$ $y$-dimensional original matrix
clocks---one component from the vector clock, one from the two-dimensional
matrix clock, and so on through one component of the $x$-dimensional matrix
clock. As we argue in Section 3, following a brief discussion of the
resource-sharing context in Section 2, this simplification leads to matrix
clocks of size $nx$, therefore considerably less complex than the $O(n^x)$-sized
original matrix clocks.

The single other attempt at reducing the complexity of a matrix clock that we
know of was given for the $x=2$ case specifically and culminated with the
introduction of two techniques [20]. The first one requires attachments of size
$O(n)$ (a considerable reduction from the original $O(n^2)$), but is only
applicable if the full asynchronism we have been assuming does not hold; it is
therefore of little interest in our context. The other technique is somewhat
closer to our own in spirit, since it aims at approximating the two-dimensional
matrix clock by retaining only $k$ of the $O(n)$ components that correspond to
each node. However, the criterion to select the components to be retained is to
choose the $k$ greatest components, which seems unrelated to our own criterion,
based as it is on the ${\it Pred}_j$ sets.

We give concluding remarks in Section 4, after a discussion of how the technique
of Section 3 can successfully solve the problem posed in Section 2.

\bigbeginsection 2. Monitoring resource-sharing computations

The resource-sharing computation we consider is one of the classical solutions
to the paradigmatic Dining Philosophers Problem (DPP) [8] in generalized form
[6]. In this case, $G$'s edge set is constructed from a given set of resources
and from subsets of that set, one for each node, indicating which resources can
be ever needed by that node. This construction places an edge between nodes $i$
and $j$ if the sets of resources ever to be needed by $i$ and $j$ have a
nonempty intersection. Notice that this construction is consonant with the
interpretation of edges as bidirectional communication channels, because it
deploys edges between every pair of nodes that may ever compete for a same
resource and must therefore be able to communicate with each other to resolve
conflicts.

In DPP, the computation carried out by a node makes it cycle endlessly through
three states, which are identified with the conditions of being idle, being in
the process of acquiring exclusive access to the resources it needs, and using
those resources for a finite period of time. While in the idle state, the node
starts acquiring exclusive access to resources when the need arises to compute
on shared resources. It is a requirement of DPP that the node must acquire
exclusive access to all the resources it shares with all its neighbors, so it
suffices for the node to acquire a token object it shares with each of its
neighbors (the ``fork,'' as it is called), each object representing all the
resources it shares with that particular neighbor. When in possession of all
forks, the node may start using the shared resources [2, 6].

The process of collecting forks from neighbors follows a protocol based
on the sending of request messages by the node that needs the forks and the
sending of the forks themselves by the nodes that have them. More than one
protocol exists, each implementing a different rule to ensure the absence of
deadlocks and lockouts during the computation. The solution we consider in this
section is based on relative priorities assigned to nodes. Another prominent
solution is also based on the assignment of relative priorities, but to the
resources instead of to the nodes [16, 23].

The priority scheme of interest to us is based on the graph-theoretic concept
of an acyclic orientation of $G$, which is an assignment of directions to the
edges of $G$ in such a way that directed cycles are not formed. Such an acyclic
orientation is then a partial order of the nodes of $G$, and is as such suitable
to indicate, given a pair of neighbors, which of the two has priority over the
other. Most of the details of this priority scheme are not relevant to our
present discussion, but before stating its essentials we do mention that the
lockout-freedom requirement leads the acyclic orientation to be changed
dynamically (so that relative priorities are never fixed), which in turn leads
to a rich dynamics in the set of all the acyclic orientations of $G$ and to
important concurrency-related issues [3--5].

Once a priority scheme is available over the set of nodes, what matters to us
is how it is used in the fork-collecting protocol. When a request for fork
arrives at node $j$ from node $i$, $j$ sends $i$ the fork they share if $j$
is either idle or is also collecting forks but does not have priority over $i$.
If $j$ is also collecting forks and has priority over $i$, or if $j$ is using
shared resources, then the sending of the fork to $i$ is postponed to until $j$
has finished using the shared resources. Note that two types of wait may happen
here. If $j$ is using shared resources when the request arrives, then the wait
is independent of $n$. If $j$ is also collecting forks, then the wait for $j$
to start using the shared resources and ultimately to send $i$ the fork is in
the worst case $n-1$ [2, 4, 6]. The reason for this is simple: $j$ is waiting
for a fork from another node, which may in turn be waiting for a fork from yet
another node, and so on. Because the priority scheme is built on acyclic
orientations of $G$, such a chain of waits does necessarily end and is $n-1$
nodes long in the worst case.

Whether such long waits occur or not is of course dependent upon the details of
each execution of the resource-sharing computation. But if they do occur, one
possibility for reducing the average wait is to increase the availability of
certain critical resources so that $G$ becomes less dense [2, 4, 5]. Perhaps
another possibility would be to fine-tune some of the characteristics of each
node's participation in the overall computation, such as the duration of its
idle period, which could be subject to a mandatory lower bound, for example.
In any event, the ability to locally detect long waits (a global property,
since it relates to the notion of time in fully asynchronous distributed
systems [2]) and identify the nodes at which the wait chains end is crucial.

To see how this relates to the formalism of Section 1, suppose our
resource-sharing computation consists of the exchange of fork-bearing messages
only (that is, request messages and all other messages involved in the
computation, such as those used to update acyclic orientations, are ignored).
For distinct nodes $i$ and $j$, and for an event $v'$ occurring at node $i$
at the reception of a fork, the set ${\it Pred}^{(x)}_j(v')$ for $x\ge 1$ is
either empty or only contains events that correspond to the sending of forks by
node $j$. Now suppose we systematically investigate the sets
${\it Pred}^{(x)}_j(v')$ for every appropriate $j$ by letting $x$ increase from
$1$ through $n-1$. Suppose also that, for each $j$, we record the first $x$ that
is found such that ${\it Pred}^{(x)}_j(v')$ contains the sending of a fork as
response to the reception of a request message without any wait for fork
collection on the part of $j$. The greatest such value to be recorded, say
$x^*$, has special significance: it means that the eventual reception of a fork
by node $i$ through $v'$ is the result of an $x^*$-long chain of waits.

If this were the only information of interest, then Lamport's clocks [15] could
be used trivially to discover it. However, taking corrective measures may
require a wider array of long wait chains to be found (not simply the longest),
as well as the nodes at which those chains end. The matrix clock that we
introduce in Section 3 is capable of conveying this information to node $i$
succinctly, so long as there exists enough redundancy in the sets
${\it Pred}^{(x)}_j(v')$ that attaching only $nX$ integers to forks suffices,
where $X$ is a threshold in the interval $[1,n-1]$ beyond which wait chains
are known not to occur, given the structure of $G$ and the initial arrangement
of priorities.\footnote{$^1$}{Discovering the value of $X$ is not necessarily a
simple task, but some empirical knowledge has already been accumulated for
modestly-sized systems [3]; in any case, in the likely event that $X$ cannot be
determined with certainty, there is always the possibility of adaptation as the
resource-sharing computation is observed on the system at hand.} It so happens
that such redundancy clearly exists: the only events in ${\it Pred}^{(x)}_j(v')$
that matter are those corresponding to the sending of forks without any wait for
fork collection. Detecting any one of them suffices, so we may as well settle
for the latest, that is, one single event from the whole set. We return to this
in Section 4.

\bigbeginsection 3. A simpler matrix clock

For $x\ge 1$, the new matrix clock we introduce is an $x\times n$ matrix. For
node $i$ at time $t_i$ (i.e., following the occurrence of $t_i$ events at node
$i$), it is denoted by $C^i(t_i)$. For $1\le y\le x$ and $1\le j\le n$,
component $C^i_{y,j}(t_i)$ of $C^i(t_i)$ is defined as follows. If $t_i=0$,
then $C^i_{y,j}(t_i)=0$, as for vector clocks and the matrix clocks of
Section 1. If $t_i\ge 1$, then we have
$$
C^i_{y,j}(t_i)=\cases{
t_i,&if $y=1$ and $j=i$;\cr
\max_{i\neq j_1\neq\cdots\neq j_y=j}
\Bigl\{{\it time}_{j_y}\Bigl({\it pred}_{j_y}\ldots{\it pred}_{j_1}
\bigl({\it event}_i(t_i)\bigr)\Bigr)\Bigr\},
&if $y>1$ or $j\neq i$.\cr
}\eqno(3)
$$
Note, first of all, that for $y=1$ this definition is equivalent to the
definition of a vector clock in (1). Thus, the first row of $C^i(t_i)$ is the
vector clock $V^i(t_i)$. For $y>1$, the definition in (3) implies, according to
the interpretation of matrix clocks that follows our definition in (2), that
$$
C^i_{y,j}(t_i)=\cases{
\max\bigl\{{\it time}_j(v)\bigm|v\in{\it Pred}^{(y)}_j(v')\bigr\},
&if ${\it Pred}^{(y)}_j(v')\neq\emptyset$;\cr
0,&if ${\it Pred}^{(y)}_j(v')=\emptyset$,\cr
}\eqno(4)
$$
where $v'$ is the $t_i$th event occurring at node $i$. What this means is that,
of all the $O(n^{y-1})$ events that may exist in ${\it Pred}^{(y)}_j(v')$, only
one (the one to have occurred latest at node $j$) makes it to the matrix clock
$C^i(t_i)$. Note also that (3) implies the equality in (4) for $y=1$ as well, so
long as $j\neq i$. In this case, ${\it Pred}^{(y)}_j(v')$, if nonempty, is the
singleton $\bigl\{{\it pred}_j(v')\bigr\}$.

Before we derive the update rules for our new matrix clock, let us pause and
examine an example. Consider Figure 1, where a computation on six nodes is
illustrated by means of the usual message diagram that forms the basis of the
relation $\prec$. Nodes are numbered $1$ through $6$, and in the figure local
time elapses from left to right independently for each node. Filled circles
represent events and the arrows connecting two events indicate messages.

\topinsert
$$
\epsfbox{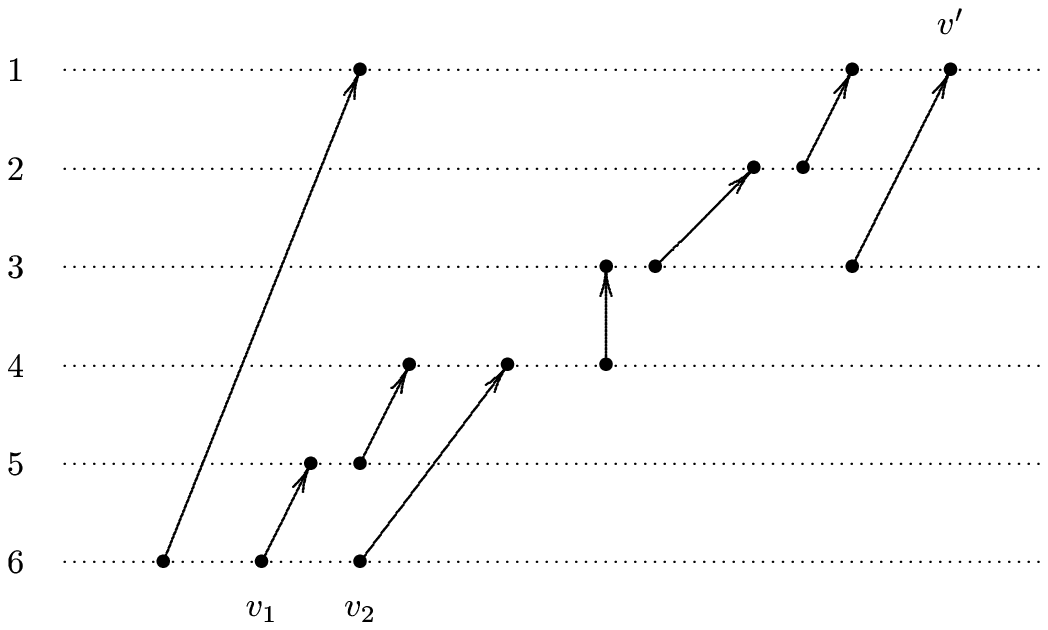}
$$
\bigskip
\centerline{{\bf Figure 1.} A computation fragment on six nodes.}
\endinsert

Three events are singled out in Figure 1, namely $v'$, $v_1$, and $v_2$. They
are related to one another in such a way that
${\it Pred}^{(4)}_6(v')=\{v_1,v_2\}$, that is, $v_1$ and $v_2$ are the
``depth-$4$'' predecessors of $v'$ at node $6$. Recalling that
${\it time}_1(v')=3$, there are two components of the four-dimensional matrix
clock $M^1(3)$ that reflect this relationship among the three events, namely
$$
M^1_{3,4,5,6}(3)={\it time}_6(v_1)=2
$$
and
$$
M^1_{2,3,4,6}(3)={\it time}_6(v_2)=3.
$$

These follow directly from (2) with $x=4$. By (3) with $y=4$, the same diagram
of Figure 1 yields
$$
\eqalign{
C^1_{4,6}(3)
&=\max\Bigl\{
{\it time}_6\bigl({\it pred}_6{\it pred}_5{\it pred}_4{\it pred}_3(v')\bigr),
{\it time}_6\bigl({\it pred}_6{\it pred}_4{\it pred}_3{\it pred}_2(v')\bigr)
\Bigr\}\cr
&=\max\bigl\{{\it time}_6(v_1),{\it time}_6(v_2)\bigr\}\cr
&=3,
}
$$
which is clearly also in consonance with (4).

Let us now look at the update rules. For $y=1$, these rules are the same as
those given for vector clocks in Section 1. For $y>1$, we know from the
definition of ${\it Pred}^{(y)}_j$ that $v\in{\it Pred}^{(y)}_j(v')$ if and only
if there exists $k\neq i$ such that an event $\bar v$ occurs at node $k$ for
which $\bar v={\it pred}_k(v')$ and $v\in{\it Pred}^{(y-1)}_j(\bar v)$. Thus,
if we let $t_k={\it time}_k(\bar v)$ and consider the matrix clocks resulting
from the occurrence of $\bar v$ and of $v'$ ($C^k(t_k)$ and $C^i(t_i)$,
respectively), then it follows from (4) that
$$
C^i_{y,j}(t_i)=\cases{
\max_{k\in K^i(t_i,y,j)}C^k_{y-1,j}(t_k),&if $K^i(t_i,y,j)\neq\emptyset$;\cr
0,&if $K^i(t_i,y,j)=\emptyset$,\cr
}\eqno(5)
$$
where $K^i(t_i,y,j)$ is the set comprising every appropriate $k$. Notice, in
(5), that $k=j$ can never occur as the maximum is taken if $y=2$: aside from
$i$, node $j$ is the only node that cannot possibly be a member of
$K^i(t_i,2,j)$, by (3).

According to (5), and recalling once again the special case of $y=1$, we are
then left with the following two rules for the evolution of our new matrix
clocks:

{\parindent=\textitemamount

\medskip
\item{$\bullet$} Upon sending a message to one of its neighbors, node $i$
attaches $C^i(t_i)$ to the message, where $t_i$ is assumed to already account
for the message that is being sent.

\medskip
\item{$\bullet$} Upon receiving a message from node $k$ with attached matrix
clock $C^k$, and assuming that $t_i$ already reflects the reception of the
message, node $i$ sets $C^i_{y,j}(t_i)$ to
$$\cases{
t_i,&if $y=1$ and $j=i$;\cr
\max\bigl\{C^i_{y,j}(t_i-1),C^k_{y,j}\bigr\},&if $y=1$ and $j\neq i$;\cr
C^i_{y,j}(t_i-1),&if $y=2$ and $j=k$;\cr
\max\bigl\{C^i_{y,j}(t_i-1),C^k_{y-1,j}\bigr\},
&if $1<y\le x$, provided $y>2$ or $j\neq k$.\cr
}
$$

}

\medskip\noindent
According to these rules, every message carries an attachment of $nx$ integers.

\bigbeginsection 4. Discussion and concluding remarks

We are now in a position to return to the problem posed in Section 2, namely the
problem of monitoring executions of the solution to DPP that employs a partial
order on $G$'s set of nodes to establish priorities. As we discussed in that
section, the overall goal is to allow nodes to detect locally, upon receiving a
fork, whether the delivery of that fork is the result of a chain of fork
deliveries that started too far back in the past. In the affirmative case, the
wait since the fork was requested will have been too long, in terms of the usual
notions of time complexity in asynchronous distributed computations.

More specifically, if $v'$ is the event corresponding to the reception of a fork
at node $i$, then the goal is for $i$ to be able to detect the existence of an
event in ${\it Pred}^{(y)}_j(v')$ that corresponds to the sending of a fork by
node $j$ either immediately upon the reception of the request for that fork or,
if node $j$ was using shared resources when the request arrived, immediately
upon finishing. Here $1\le y\le X$ and $j$ is any node, provided $y=1$ and $j=i$
do not occur in conjunction. The value of $X$ is such that $1\le X\le n-1$, and
is chosen as a bound to reflect the maximum possible chain of waits. As a
consequence, the sets ${\it Pred}^{(y)}_j(v')$ must include fork-related events
only.

The new matrix clocks introduced in Section 3 can be used for this detection
with only minimal adaptation. The key ingredients are:

{\parindent=\textitemamount

\medskip
\item{$\bullet$} The only messages sent by node $i$ to be tagged with the matrix
clock $C^i$ are forks. If the sending of a fork by node $i$ does not depend on
further fork collection by $i$, then every component of $C^i$ other than
$C^i_{1,i}$ is reset to zero before it is attached to the fork. Matrix clocks
are $X\times n$, and all further handling of them follows the rules given in
Section 3.

\medskip
\item{$\bullet$} Upon receiving a fork with attached matrix clock, and having
updated $C^i$ accordingly, node $i$ looks for components of $C^i$ that contain
nonzero values. If $C^i_{y,j}$ is one such component for $y>1$ or $j\neq i$,
then a wait chain of length $y$ that ends at node $j$ has been discovered and
can be checked against a certain threshold $X'<X$ representing the maximum
allowable chain length. If $C^i$ contains zero components at all positions but
$(1,i)$, then it is certain that the value of $X$ has to be revised, as clearly
a wait chain exists whose length surpasses $X$. A greater value for $X$ is then
needed.

}

\medskip\noindent
This strategy reflects the general method of Section 3 when applied to events
that relate to the flow of forks only. Whenever the request for a fork is
received and the fork can be sent without the need for any forks to be received
by the node in question, say node $j$, zeroes get inserted into the matrix clock
at all positions but $(1,j)$ and are sent along with the fork. The sending of
this fork may unleash the sending of forks by other nodes in a chain of events,
and along the way the original value of $C^j_{1,j}$ may never be reset to zero,
reflecting the increasing length of the wait chain rooted at node $j$. The
reception of a fork whose matrix clock has such a nonzero component beyond row
$X'$ is then an indication that such reception is part of chains whose lengths
are considered too long.

It is now worth returning to Figure 1 in order to view its diagram as
representing the fork-bearing messages in a fragment of a DPP computation of
the type we have been considering. For such, consider Figure 2, where two
six-node graphs are depicted. Figure 2(a) shows the graph $G$ corresponding to
a certain resource-sharing pattern for the six nodes; it shows an acyclic
orientation of the edges of $G$ as well, indicating the initial arrangement
of priorities. In a situation of heavy demand for resources by all six nodes,
only node $6$ can acquire all the (three) forks it needs and proceed. All others
must wait along the length-$5$ wait chain  shown in Figure 2(b): node $1$ for
node $2$, $2$ for $3$, and so on through node $6$.

\topinsert
$$
\epsfbox{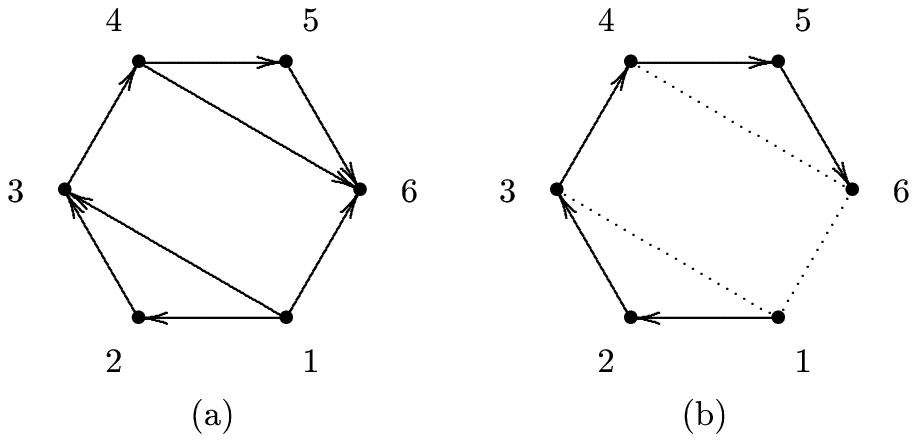}
$$
\bigskip
\centerline{{\bf Figure 2.} $G$ oriented acyclically (a) and a possible wait
chain (b).}
\endinsert

Eventually, it must happen that node $6$ sends out its three forks. If we
assume that the corresponding request messages arrive when node $6$ already
holds all three forks, then upon sending them out all components of $C^6$ are
sent as zeroes, except for $C^6_{1,6}$, sent as $1$, $2$, or $3$, depending on
the destination. For $X=5$, at the occurrence of $v'$ node $1$ updates its
matrix clock in such a way that $C^1_{5,6}=2$, which characterizes the
length-$5$ wait chain ending at node $6$.

In summary, we have in this paper introduced a novel notion of matrix clocks.
Similarly to the original matrix clocks [21, 24], whose definition appears in
(2), our matrix clock has the potential of reflecting causal dependencies in
the flow of messages that stretch as far as depth $x$ into the past. Unlike
those original matrix clocks, however, ours increases with $x$ as $nx$, while
in the original case the growth is according to an $O(n^x)$ function, that is,
exponentially.

We have illustrated the applicability of the new matrix clocks with an example
from the area of resource-sharing problems. What we have demonstrated is a means
of collecting information locally during the resource-sharing computation so
that exceedingly long global waits can be detected, possibly indicating the need
for overall system re-structuring, as described in Section 2.

\beginsection Acknowledgments

The authors acknowledge partial support from CNPq, CAPES, the PRONEX initiative
of Brazil's MCT under contract 41.96.0857.00, and a FAPERJ BBP grant.

\bigbeginsection References

{\frenchspacing

\medskip
\item{1.} M. Ahuja, T. Carlson, and A. Gahlot,
``Passive-space and time view: vector clocks for achieving higher performance,
program correction, and distributed computing,''
{\it IEEE Trans. on Software Engineering\/} {\bf 19} (1993),
845--855.

\medskip
\item{2.} V. C. Barbosa,
{\it An Introduction to Distributed Algorithms},
The MIT Press, Cambridge, MA, 1996.

\medskip
\item{3.} V. C. Barbosa,
{\it An Atlas of Edge-Reversal Dynamics},
Chapman \& Hall/CRC, London, UK, 2000.

\medskip
\item{4.} V. C. Barbosa,
``The combinatorics of resource sharing,''
in R. Corr\^ea {\it et alii\/} (Eds.),
{\it Models for Parallel and Distributed Computation:
Theory, Algorithmic Techniques and Applications}, 27--52,
Kluwer Academic Publishers, Dordrecht, The Netherlands, 2002.

\medskip
\item{5.} V. C. Barbosa and E. Gafni,
``Concurrency in heavily loaded neighborhood-constrained systems,''
{\it ACM Trans. on Programming Languages and Systems\/} {\bf 11} (1989),
562--584.

\medskip
\item{6.} K. M. Chandy and J. Misra,
``The drinking philosophers problem,''
{\it ACM Trans. on Programming Languages and Systems\/} {\bf 6} (1984),
632--646.

\medskip
\item{7.} B. Charron-Bost,
``Concerning the size of logical clocks in distributed systems,''
{\it Information Processing Letters\/} {\bf 39} (1991),
11--16.

\medskip
\item{8.} E. W. Dijkstra,
``Hierarchical ordering of sequential processes,''
{\it Acta Informatica\/} {\bf 1} (1971),
115--138.

\medskip
\item{9.} R. P. Dilworth,
``A decomposition theorem for partially ordered sets,''
{\it Annals of Mathematics\/} {\bf 51} (1950),
161--165.

\medskip
\item{10.} L. M. A. Drummond and V. C. Barbosa,
``Distributed breakpoint detection in message-passing programs,''
{\it J. of Parallel and Distributed Computing\/} {\bf 39} (1996), 153--167.

\medskip
\item{11.} C. J. Fidge,
``Timestamps in message-passing systems that preserve partial ordering,''
{\it Proc. of the 11th Australian Computer Science Conference}, 56--66, 1988.

\medskip
\item{12.} C. J. Fidge,
``Logical time in distributed computing systems,''
{\it IEEE Computer\/} {\bf 24} (1991),
28--33.

\medskip
\item{13.} C. J. Fidge,
``Fundamentals of distributed system observation,''
{\it IEEE Software\/} {\bf 13} (1996),
77--83.

\medskip
\item{14.} V. K. Garg,
{\it Principles of Distributed Systems},
Kluwer Academic Publishers, Boston, MA, 1996.

\medskip
\item{15.} L. Lamport,
``Time, clocks, and the ordering of events in a distributed system,''
{\it Comm. of the ACM\/} {\bf 21} (1978),
558--565.

\medskip
\item{16.} N. A. Lynch,
``Upper bounds for static resource allocation in a distributed system,''
{\it J. of Computer and System Sciences\/}
{\bf 23} (1981), 254--278.

\medskip
\item{17.} F. Mattern,
``Virtual time and global states in distributed systems,''
in M. Cosnard {\it et alii\/} (Eds.),
{\it Parallel and Distributed Algorithms: Proc. of the Int. Workshop on Parallel
and Distributed Algorithms}, 215--226, North-Holland, Amsterdam,
The Netherlands, 1989.

\medskip
\item{18.} M. Raynal,
``Illustrating the use of vector clocks in property detection: an example and
a counter-example,''
in P. Amestoy {\it et alii\/} (Eds.),
{\it Euro-Par'99---Parallel Processing}, 806--814,
Lecture Notes in Computer Science 1685, Springer-Verlag, Berlin, Germany, 1999.

\medskip
\item{19.} L. E. T. Rodrigues and P. Ver\'\i ssimo,
``Causal separators for large-scale multicast communication,''
{\it Proc. of the 15th Int. Conf. on Distributed Computing Systems},
83--91, 1995.

\medskip
\item{20.} F. Ruget,
``Cheaper matrix clocks,''
in G. Tel and P. Vit\'anyi (Eds.),
{\it Distributed Algorithms: Proc. of the 8th Int. Workshop on Distributed
Algorithms}, 355--369,
Lecture Notes in Computer Science 857, Springer-Verlag, Berlin, Germany, 1994.

\medskip
\item{21.} S. K. Sarin and L. Lynch,
``Discarding obsolete information in a replicated database system,''
{\it IEEE Trans. on Software Engineering\/} {\bf SE-13} (1987), 39--46.

\medskip
\item{22.} M. Singhal and A. Kshemkalyani.
``An efficient implementation of vector clocks,''
{\it Information Processing Letters\/} {\bf 43} (1992),
47--52.

\medskip
\item{23.} J. L. Welch and N. A. Lynch,
``A modular drinking philosophers algorithm,''
{\it Distributed Computing\/} {\bf 6} (1993),
233--244.

\medskip
\item{24.} G. T. J. Wuu and A. J. Bernstein,
``Efficient solutions to the replicated log and dictionary problems,''
{\it Proc. of the 3rd Annual ACM Symposium on Principles of Distributed
Computing}, 233--242, 1984.

}

\bye